\documentclass[11pt]{article}
\usepackage{fullpage,epsf,amssymb,amsthm,amsfonts,amsmath,latexsym,array,graphicx,extarrows,mathtools,mathstyle,mathrsfs,slashed,subcaption,amsbsy,csquotes,accents}
\usepackage[normalem]{ulem}
\usepackage{authblk}
\usepackage{cite}
\usepackage{color}

\usepackage{float}

\usepackage{hyperref}

\usepackage[export]{adjustbox}

\numberwithin{equation}{section}

\title{\bf{Spectral representation of the shear viscosity for local scalar QFTs at finite temperature}}

\author[1]{Peter Lowdon\thanks{lowdon@itp.uni-frankfurt.de}}
\author[2]{Ralf-Arno Tripolt\thanks{ralf.tripolt@uni-graz.at}}
\author[3,4]{Jan M.~Pawlowski\thanks{j.pawlowski@thphys.uni-heidelberg.de}}
\author[1,5]{Dirk H.~Rischke\thanks{drischke@itp.uni-frankfurt.de}}

\affil[1]{Institut f\"{u}r Theoretische Physik, Johann Wolfgang Goethe-Universit\"{a}t, Max-von-Laue-Str. 1,  60438 Frankfurt am Main, Germany}
\affil[2]{Institut f\"{u}r Physik, Karl-Franzens-Universit\"{a}t Graz, NAWI Graz, Universit\"{a}tsplatz 5, 8010 Graz, Austria}
\affil[3]{Institut f\"{u}r Theoretische Physik, Universit\"{a}t Heidelberg, Philosophenweg 16, 69120 Heidelberg, Germany}
\affil[4]{ExtreMe Matter Institute EMMI, GSI, Planckstr. 1, 64291 Darmstadt, Germany}
\affil[5]{Helmholtz Research Academy Hesse for FAIR, Campus Riedberg, Max-von-Laue-Str. 12, 60438 Frankfurt am Main, Germany}

\date{}
\begin{document}

{\let\newpage\relax\maketitle}
\setcounter{page}{1}
\pagestyle{plain}

\begin{abstract}
\noindent
In local scalar quantum field theories (QFTs) at finite temperature correlation functions are known to satisfy certain non-perturbative constraints, which for two-point functions in particular implies the existence of a generalisation of the standard K\"{a}ll\'{e}n-Lehmann representation. In this work, we use these constraints in order to derive a spectral representation for the shear viscosity arising from the thermal asymptotic states, $\eta_{0}$. As an example, we calculate $\eta_{0}$ in $\phi^{4}$ theory, establishing its leading behaviour in the small and large coupling regimes. 
\end{abstract}

\newpage

\section{Introduction}
\label{intro}

Determining the properties of quantum field theories (QFTs) at finite temperature is essential for describing many physical phenomena. Whilst progress has been made, particularly for systems in or approaching equilibrium, the full extent to which QFTs are modified by the presence of a thermal background medium remains largely unknown. In order to fully understand these effects one ultimately requires a framework that does not depend on the specific coupling regime. At zero temperature, a successful such framework was developed by defining QFTs using a series of physically motivated axioms~\cite{Streater:1989vi,Haag:1992hx,Bogolyubov:1990kw}. The advantage of this approach is that it allows non-perturbative characteristics to be derived in a purely analytic manner, and has led to numerous important insights such as the \textit{CPT} theorem, collision theory, and the rigorous connection of Minkowski and Euclidean QFTs. A natural question is whether this framework can be extended beyond zero temperature. In Refs.~\cite{Bros:1992ey,Bros:1995he,Bros:1998ua,Bros:1996mw} the first important steps were taken to demonstrate that for Hermitian scalar fields this is indeed possible, but crucially this requires a modification of the standard axioms. \\

\noindent
An essential difference in the formulation of finite temperature QFT proposed in Refs.~\cite{Bros:1992ey,Bros:1995he,Bros:1998ua,Bros:1996mw} is the existence of a thermal equilibrium state $|\Omega_{\beta}\rangle$ at temperature $T=1/\beta$. Unlike the vacuum state $|0\rangle$ at vanishing temperature, the thermal state $|\Omega_{\beta}\rangle$ determines a privileged reference frame, and hence full Lorentz symmetry cannot exist. This is reflected in the assumption that the corresponding quantised fields $\phi(x)$ no longer transform covariantly under a unitary representation of the full Poincar\'{e} group. Moreover, whilst $|0\rangle$ is a unique Poincar\'{e} invariant state, $|\Omega_{\beta}\rangle$ constitutes a thermal superposition, and only remains invariant under the subgroup of spacetime translations and rotations. Temporal invariance is implied by the fact that $|\Omega_{\beta}\rangle$ is an equilibrium state, and hence stationary, whereas spatial translational and rotational invariance is a choice that assumes the thermal system to be both homogeneous and isotropic~\cite{Bros:1992ey}. Despite these differences, there are nevertheless several key assumptions that remain unchanged, including the distributional nature of the fields\footnote{$\phi(x)$ are defined to be operator-valued (tempered) distributions, and hence only the smeared fields $\int d^{4}x \, f(x)\phi(x)$ have meaning as well-defined operators~\cite{Streater:1989vi,Haag:1992hx}.}, their locality\footnote{Locality requires that the fields commute with one another for space-like separations.}, and the fact that the states in the theory are constructed by acting with the fields on the background state, in this case $|\Omega_{\beta}\rangle$. It subsequently follows that the correlation functions of the fields $\langle \Omega_{\beta}|\phi(x_{1})\cdots\phi(x_{n})|\Omega_{\beta}\rangle$, the so-called \textit{thermal correlation functions}, encode all of the dynamical properties of the QFT, just like in the vacuum theory~\cite{Bros:1996mw}. Determining the properties of these objects is therefore central to understanding the characteristics of finite temperature QFTs. \\

\noindent
At high temperatures, collective relativistic particle systems display fluid-like behaviour, and hence hydrodynamical quantities such as transport coefficients are important observables. A coefficient of particular relevance is the shear viscosity $\eta$, which describes the resistiveness to sheared flow. Over recent years, a significant theoretical effort has been dedicated towards calculating $\eta$ in different models using a variety of methods. Perturbative calculations of $\eta$ have been performed in many instances, including in physical theories such as quantum chromodynamics (QCD)~\cite{Arnold:2000dr,Arnold:2003zc,Ghiglieri:2018dib}. However, even for the simplest scalar theories difficulties arise due to the appearance of infrared divergences, which leads to a worsening of the perturbative convergence~\cite{Jeon:1994if,Jeon:1995zm}. In essence, this stems from non-perturbative corrections due to the interactions with the background medium. In order to circumvent these issues, non-perturbative techniques like lattice QFT~\cite{Nakamura:2004sy,Meyer:2007ic,Meyer:2011gj,Astrakhantsev:2017nrs,Borsanyi:2018srz}, together with functional methods such as the functional renormalisation group (FRG) and Dyson-Schwinger equations (DSEs)~\cite{Haas:2013hpa,Christiansen:2014ypa}, have been applied in order to calculate $\eta$ directly. A problem with Euclidean-based techniques like lattice QFT is that non-unique numerical inversions must be performed in order to reconstruct $\eta$ from the Euclidean data. Although functional methods can in principle avoid this issue, model dependent information is still necessary to establish the form of $\eta$. In light of these theoretical difficulties there is clearly a strong motivation to better understand the analytic structure of $\eta$. The goal of this work will be to use the non-perturbative constraints of local QFT to provide new insights.  \\

\noindent
The remainder of this paper is structured as follows: in Sec.~\ref{sec:outline} we outline the general non-perturbative constraints imposed on thermal correlation functions, in particular the two-point function of real scalar fields; in Sec.~\ref{shear_sec} we use these constraints to derive a spectral representation for the shear viscosity arising from the thermal asymptotic states $\eta_{0}$, which in Sec.~\ref{shear_part} we apply in order to calculate an explicit expression for $\eta_{0}$ in $\phi^{4}$ theory. Finally, in Sec.~\ref{concl} we summarise our key findings.

\section{Thermal correlation functions in local QFT}
\label{sec:outline}

As outlined in Sec.~\ref{intro}, in Refs.~\cite{Bros:1992ey,Bros:1995he,Bros:1998ua,Bros:1996mw} it was demonstrated that the standard assumptions of zero temperature local QFT can be adapted in order to describe systems in thermal equilibrium. It turns out that the conditions of locality, translational invariance, and thermal equilibrium impose particularly significant analytic constraints. In this section we will analyse the consequences of these constraints for the structure of the thermal correlation functions.

\subsection{General constraints}
\label{local_assum}

Since the quantised fields $\phi(x)$ are defined to be operator-valued (tempered) distributions, it immediately follows that the thermal correlation functions $\langle \Omega_{\beta}|\phi(x_{1})\cdots\phi(x_{n})|\Omega_{\beta}\rangle$ are distributions, just like at vanishing temperature. For a real scalar theory to be local this requires that $\left[\phi(x),\phi(y)\right]=0$ for $(x-y)^{2}<0$. Because locality is an operator identity it must hold for all states, and is therefore independent of the representation of the system. This implies that the corresponding constraints on the thermal correlation functions are identical to those in the vacuum theory. In particular, it follows that
\begin{align}
\langle \Omega_{\beta}|\phi(x_{1})\cdots\phi(x_{k})\phi(x_{k+1})\cdots \phi(x_{n})|\Omega_{\beta}\rangle = \langle \Omega_{\beta}|\phi(x_{1})\cdots\phi(x_{k+1})\phi(x_{k})\cdots \phi(x_{n})|\Omega_{\beta}\rangle,
\label{locality}
\end{align} 
for $(x_{k}-x_{k+1})^{2}<0$. Similarly, due to the assumption that $|\Omega_{\beta}\rangle$ is invariant under spacetime translations and the fields transform covariantly under the action of this symmetry, one finds that ($\forall a \in \mathbb{R}^{4}$)
\begin{align}
\langle \Omega_{\beta}|\phi(x_{1})\phi(x_{2})\cdots \phi(x_{n})|\Omega_{\beta}\rangle = \langle \Omega_{\beta}|\phi(x_{1}+a)\phi(x_{2}+a)\cdots \phi(x_{n}+a)|\Omega_{\beta}\rangle. 
\label{translation} 
\end{align} 
So far, the correlation function constraints are identical to those in the vacuum theory, except that $|\Omega_{\beta}\rangle$ is no longer invariant under Lorentz symmetry. The fundamental differences arise from the fact that $|\Omega_{\beta}\rangle$ is an equilibrium state. The physical requirement of thermal equilibrium is captured via the Kubo-Martin-Schwinger (KMS) condition~\cite{Haag:1967sg}:
\begin{align}
&\langle \Omega_{\beta}|\phi(x_{1})\cdots \phi(x_{k})\phi(x_{k+1})\cdots \phi(x_{n})|\Omega_{\beta}\rangle  \nonumber \\
& \quad\quad\quad\quad\quad\quad\quad\quad\quad\quad = \langle \Omega_{\beta}|\phi(x_{k+1})\cdots \phi(x_{n}) \phi(x_{1}+i(\beta,\vec{0}))\cdots \phi(x_{k}+i(\beta,\vec{0}) )|\Omega_{\beta}\rangle,
\label{KMS}
\end{align}
which holds $\forall k\in \{1,\dots,n-1\}$ and all thermal $n$-point functions\footnote{More concretely, the KMS condition implies that $\langle \Omega_{\beta}|\phi(x_{k+1})\cdots \phi(x_{n}) \phi(x_{1}+z)\cdots \phi(x_{k}+z )|\Omega_{\beta}\rangle$ defines an analytic continuation of the thermal $n$-point function in the variable $z \in \mathbb{C}^{4}$, which is holomorphic in the region $0 < \text{Im}z_{0} < \beta$, and has boundary values $\langle \Omega_{\beta}|\phi(x_{k+1})\cdots\phi(x_{n})\phi(x_{1})\cdots\phi(x_{k})|\Omega_{\beta}\rangle$ and $\langle \Omega_{\beta}|\phi(x_{1})\cdots\phi(x_{n})|\Omega_{\beta}\rangle$ at the two endpoints, respectively.}. It was further established in Ref.~\cite{Bros:1998ua} that the condition in Eq.~\eqref{KMS} can be extended in a Lorentz covariant manner, and doing so enables the constraint of thermal equilibrium to be defined for arbitrary observers. This is referred to as the relativistic KMS condition\footnote{This extension implies that $\langle \Omega_{\beta}|\phi(x_{k+1})\cdots \phi(x_{n}) \phi(x_{1}+z)\cdots \phi(x_{k}+z )|\Omega_{\beta}\rangle$ is holomorphic in the larger region $|\text{Im}z| < \text{Im}z_{0} < \beta - |\text{Im}z|$, which reduces to the standard condition for the specific point $z=i(\beta,\vec{0})$~\cite{Bros:1995he}.}~\cite{Bros:1998ua}. Throughout the remainder of this work we will take the KMS condition to mean the full relativistic constraint. \\

\noindent
The thermal two-point functions are of particular importance for understanding the analytic structure of the shear viscosity. In the next section we will outline the constraints imposed on these objects due to the non-perturbative conditions in Eqs.~\eqref{locality}-\eqref{KMS}.

\subsection{The thermal two-point function}

In the particular case of the thermal two-point function $\langle \Omega_{\beta}| \phi(x)\phi(y) |\Omega_{\beta} \rangle$, translational invariance [Eq.~\eqref{translation}] implies that the correlation function depends only on the variable $x-y$, and hence one can define $\mathcal{W}_{\beta}(x-y) = \langle \Omega_{\beta}| \phi(x)\phi(y) |\Omega_{\beta} \rangle$. If one then takes the Fourier transform of the KMS condition [Eq.~\eqref{KMS}] with respect to $x-y$ for $n=2$, it follows that 
\begin{align}
\widetilde{\mathcal{W}}_{\beta}(p) = e^{\beta p_{0}}\widetilde{\mathcal{W}}_{\beta}(-p),
\label{KMS_p}
\end{align}
which when combined with the definition of the thermal two-point commutator $C_{\beta}(x-y) =\langle \Omega_{\beta}| \left[\phi(x),\phi(y)\right]|\Omega_{\beta} \rangle$ immediately implies the well-known condition
\begin{align}
\widetilde{\mathcal{W}}_{\beta}(p) = \frac{\widetilde{C}_{\beta}(p)}{1-e^{-\beta p_{0}}}.
\label{W_rep}
\end{align}  
Equation~\eqref{W_rep} demonstrates that similarly to the vacuum theory, where 
\begin{align}
\widetilde{\mathcal{W}}_{\text{vac}}(p) = \theta(p_{0}) \, \widetilde{C}_{\text{vac}}(p),
\label{W_vac}
\end{align}
the thermal two-point function can be uniquely recovered from the commutator. As one would expect, in the zero temperature limit ($\beta\rightarrow \infty$) Eq.~\eqref{W_rep} approaches the vacuum theory constraint in Eq.~\eqref{W_vac}, and is therefore only defined for non-negative energies. In local formulations of QFT this is referred to as the \textit{spectral condition}~\cite{Streater:1989vi,Haag:1992hx}. For non-vanishing temperatures, Eq.~\eqref{W_rep} implies that $\widetilde{\mathcal{W}}_{\beta}(p)$ can in general have contributions for $p_{0}<0$, but that these will decay exponentially for large values of $|p_{0}|$. From a physical perspective, this reflects the possibility of extracting energy from the background medium, which is thermodynamically suppressed as the temperature decreases~\cite{Bros:1996mw}. \\

\noindent
The conditions outlined so far in this section are well-known features of finite temperature QFT, and have been instrumental in forming the conventional treatment of this subject~\cite{Kapusta:2006pm,Bellac:2011kqa}. In particular, these conditions highlight the important role of the thermal commutator $\widetilde{C}_{\beta}(p)$, or \textit{spectral function} as it is commonly known, in determining the characteristics of these theories. In Ref.~\cite{Bros:1992ey} it was first pointed out that locality actually imposes significant additional constraints on the structure of $\widetilde{C}_{\beta}(p)$. Since locality requires that the position space commutator must vanish at space-like points, this implies that the momentum space commutator can be written in the following general manner~\cite{Bros:1992ey}: 
\begin{align}
\widetilde{C}_{\beta}(p_{0},\vec{p}) = \int_{0}^{\infty} \! ds \int \! \frac{d^{3}\vec{u}}{(2\pi)^{2}} \ \epsilon(p_{0}) \, \delta\!\left(p^{2}_{0} - (\vec{p}-\vec{u})^{2} - s \right) \widetilde{D}_{\beta}(\vec{u},s),
\label{spec_rep}
\end{align}
where $\epsilon(p_{0})$ is the sign function. As the temperature dependence is contained entirely within $\widetilde{D}_{\beta}(\vec{u},s)$, this quantity uniquely describes the effects of the thermal background medium. For vanishing temperature, the Poincar\'{e} covariance of the fields is restored, and $|\Omega_{\beta} \rangle$ approaches the vacuum state $|0\rangle$. In this limit, one finds that
\begin{align}
\widetilde{D}_{\beta}(\vec{u},s) \ \xlongrightarrow{\beta\rightarrow \infty}{} \ (2\pi)^{3}\delta^{3}(\vec{u})\, \rho(s),
\end{align}   
which after substitution into Eq.~\eqref{spec_rep} implies
\begin{align}
\widetilde{C}_{\beta}(p_{0},\vec{p}) \  \xlongrightarrow{\beta\rightarrow \infty}{} \ 2\pi \, \epsilon(p_{0}) \! \int_{0}^{\infty} \! ds \ \delta\!\left(p^{2} - s \right)\rho(s).
\label{KL_rep}
\end{align}
Since Eq.~\eqref{KL_rep} is the standard K\"{a}ll\'{e}n-Lehmann representation~\cite{Kallen:1952zz,Lehmann:1954xi}, with $\rho(s)$ the zero temperature spectral density\footnote{For example, $\rho(s)=\delta(s-m^{2})$ in a free scalar theory with mass $m$.}, Eq.~\eqref{spec_rep} therefore corresponds to the finite temperature generalisation of this representation. With this in mind, we refer to $\widetilde{D}_{\beta}(\vec{u},s)$ throughout as the \textit{thermal spectral density}. From Eq.~\eqref{spec_rep} one can explicitly see that $\widetilde{C}_{\beta}(p_{0},\vec{p})$ is anti-symmetric with respect to $p_{0}$. This follows from the overall anti-symmetry property: $\widetilde{C}_{\beta}(p)= -\widetilde{C}_{\beta}(-p)$, due to the definition of the commutator, and the condition: $\widetilde{C}_{\beta}(p_{0},\vec{p})=\widetilde{C}_{\beta}(p_{0},-\vec{p})$, which is implied by the rotational invariance of the background state. Because of the $\vec{p}$-reflectional symmetry it follows from Eq.~\eqref{spec_rep} that the thermal spectral density must satisfy
\begin{align}
\widetilde{D}_{\beta}(\vec{u},s)= \widetilde{D}_{\beta}(-\vec{u},s),
\label{D_symmetry}
\end{align}
and hence depends only on $|\vec{u}|$. For the remainder of this paper we will write $\widetilde{D}_{\beta}(\vec{u},s)$, but understand this to implicitly depend on the one-dimensional variable $|\vec{u}|$.  \\ 

\noindent
Despite the fact that Eq.~\eqref{spec_rep} imposes significant constraints on $\widetilde{C}_{\beta}(p)$, and hence on the characteristics of finite temperature QFTs as a whole, this thermal spectral representation has largely been overlooked in the literature. In the next sections we will demonstrate that this representation has important implications for the properties of particles moving within a thermal medium, and in particular on the structure of the shear viscosity.

\section{Analytic structure of the shear viscosity}
\label{shear_sec}

As with any observable in QFT, the behaviour of the shear viscosity is fixed by the correlation functions in the theory. In this section, we will use the model independent constraints outlined in Sec.~\ref{sec:outline} to derive a non-perturbative spectral representation for the shear viscosity arising from the thermal asymptotic states $\eta_{0}$, and discuss the essential role played by the thermal spectral density and its corresponding analytic properties. \\

\noindent
For a local scalar QFT at finite temperature with energy-momentum tensor $T^{\mu\nu}$, the shear viscosity $\eta$ can be calculated from the Kubo relation~\cite{Kubo:1957mj} 
\begin{align}
\eta = \frac{1}{20} \lim_{p_{0} \rightarrow 0}\frac{d \rho_{\pi\pi}}{d p_{0}},
\label{shear}
\end{align} 
where $\rho_{\pi\pi}(p_{0}) = \widetilde{C}_{\pi\pi}(p_{0},\vec{p}=0)$, and $\widetilde{C}_{\pi\pi}(p) = \mathcal{F}\left[\langle \Omega_{\beta}|  \left[\pi^{ij}(x),\pi_{ij}(y)\right]|\Omega_{\beta}\rangle\right](p)$, with $\pi^{ij} = T^{ij} - \frac{1}{3}g^{ij}T^{k}_{k}$ the spatial traceless component of the energy-momentum tensor. Independently of the specific form of the interactions, $\pi^{ij}$ is given by  
\begin{align}
\pi^{ij}= (\partial^{i}\phi)(\partial^{j}\phi) - \frac{1}{3}g^{ij}(\partial^{k}\phi)(\partial_{k}\phi).
\label{pi}
\end{align}
For the purposes of this study we are interested in calculating the shear viscosity arising from the thermal asymptotic states $\eta_{0}$, and hence only the contributions of the thermal correlation functions at asymptotic times play a role. In Ref.~\cite{Bros:2001zs}, it was demonstrated that in the limit of asymptotic temporal separations the thermal $n$-point functions decompose into products of two-point functions. By applying Eq.~\eqref{W_rep}, together with a point-splitting regularisation to make sense of the field products in Eq.~\eqref{pi}, it ultimately follows that the contribution of the asymptotic states to $\rho_{\pi\pi}(p_{0})$ can be expressed in the form
\begin{align}
\rho_{\pi\pi}(p_{0}) &=  \sinh\left(\frac{\beta}{2}p_{0}\right)\int \frac{d^{3}\vec{q}}{(2\pi)^{4}} \frac{2}{3}|\vec{q}|^{4} \int_{-\infty}^{\infty} dq_{0} \, \frac{\widetilde{C}_{\beta}(q_{0},\vec{q})\, \widetilde{C}_{\beta}(p_{0}-q_{0},\vec{q})}{\sinh\left(\frac{\beta}{2}q_{0}\right)   \sinh\left(\frac{\beta}{2}(p_{0}-q_{0})\right)},
\label{symmetry}
\end{align}
where $\widetilde{C}_{\beta}$ is the thermal (two-point) commutator. Equation~\eqref{symmetry} coincides in structure with the lowest order perturbative calculation of $\rho_{\pi\pi}$~\cite{Jeon:1992kk}, although the representation in Eq.~\eqref{symmetry} is non-perturbative. This occurs because the large-time behaviour of the correlation functions has a quasi-free structure, and hence connected components are suppressed. In analyses of the shear viscosity in the literature a variety of different methods are adopted in order to either explicitly calculate, or model the form of $\widetilde{C}_{\beta}$. Once the dependence of $\rho_{\pi\pi}$ on $\widetilde{C}_{\beta}$ is known, the Kubo relation in Eq.~\eqref{shear} can then be applied. However, in general these analyses do not take into account the additional constraints imposed on $\widetilde{C}_{\beta}$ and $\rho_{\pi\pi}$ by locality, namely the thermal spectral representation in Eq.~\eqref{spec_rep}. As we will now demonstrate, applying Eq.~\eqref{spec_rep} one can derive a spectral representation for $\eta_{0}$ and, in doing so, explicitly establish how the model-dependence of thermal scalar QFTs affects the behaviour of $\eta_{0}$.  \\

\noindent
After substituting Eq.~\eqref{spec_rep} into Eq.~\eqref{symmetry}, the $q_{0}$ convolution takes the form 
\begin{align}
\! \int_{-\infty}^{\infty} \! dq_{0} \, \frac{\widetilde{C}_{\beta}(q_{0},\vec{q})\, \widetilde{C}_{\beta}(p_{0}-q_{0},\vec{q})}{\sinh\left(\!\frac{\beta}{2}q_{0} \right)   \sinh\left(\!\frac{\beta}{2}(p_{0}-q_{0})\right)} &= \int_{0}^{\infty} \!\! ds \int_{0}^{\infty} \!\! dt \! \int \! \frac{d^{3}\vec{u}}{(2\pi)^{2}2 E_{u}}\frac{d^{3}\vec{v}}{(2\pi)^{2}2 E_{v}}  \frac{\widetilde{D}_{\beta}(\vec{u},s) \, \widetilde{D}_{\beta}(\vec{v},t)}{\sinh\left(\frac{\beta}{2}E_{u}\right)  \sinh\left(\frac{\beta}{2}E_{v}\right)} \nonumber \\[0.5em]
&  \times \Big[ \delta\left(p_{0} - E_{u}-E_{v} \right) + 2\delta\left(p_{0} -E_{u}+ E_{v} \right)  + \delta\left(p_{0} + E_{u}+ E_{v} \right)  \Big],
\label{conv}
\end{align}
where $E_{u}= \sqrt{(\vec{q}-\vec{u})^{2}+s}$, and $E_{v}= \sqrt{(\vec{q}-\vec{v})^{2}+t}$. For simplicity, we calculate the contributions to $\rho_{\pi\pi}(p_{0})$ from each of the three delta components in Eq.~\eqref{conv} separately, defining
\begin{align}
\rho_{\pi\pi}(p_{0})=\rho_{\pi\pi}^{(1)}(p_{0})+\rho_{\pi\pi}^{(2)}(p_{0})+\rho_{\pi\pi}^{(3)}(p_{0}),
\end{align}
where the numbering corresponds to the ordering in Eq.~\eqref{conv}. Since the thermal spectral density $\widetilde{D}_{\beta}(\vec{u},s)$ depends only on $|\vec{u}|$ and $s$, one can explicitly perform the angular integrals in Eq.~\eqref{conv}. Doing so for the first delta component, and applying Eq.~\eqref{symmetry}, gives
\begin{align}
\rho_{\pi\pi}^{(1)}(p_{0}) &= \int_{0}^{\infty} \!\! ds \int_{0}^{\infty} \!\! dt \int_{0}^{\infty} d|\vec{q}|\,  \frac{|\vec{q}|^{4}}{48\pi^{5}\beta}   \int_{0}^{\infty} d|\vec{u}| \int_{0}^{\infty} d|\vec{v}| \ |\vec{u}||\vec{v}|\,\widetilde{D}_{\beta}(\vec{u},s) \, \widetilde{D}_{\beta}(\vec{v},t)  \nonumber \\[0.5em]
& \quad\quad\quad\quad\quad\quad  \Bigg\{ \theta \!\left(\mathcal{E}_{u}^{+} + \mathcal{E}_{v}^{-} - p_{0} \right)   \ln\!\left[ \frac{\sinh \left(\frac{\beta}{2}(\mathcal{E}_{u}^{+} - p_{0})  \right)\sinh \left(\frac{\beta}{2}(\mathcal{E}_{v}^{-} - p_{0})  \right)   }{\sinh \left(\frac{\beta}{2}\mathcal{E}_{u}^{+}  \right)\sinh \left(\frac{\beta}{2}\mathcal{E}_{v}^{-}\right) }  \right]   \nonumber \\[0.5em]
&  \quad\quad\quad\quad\quad\quad\quad + \theta \!\left( \mathcal{E}_{u}^{-} + \mathcal{E}_{v}^{+} - p_{0} \right)  \ln\!\left[ \frac{\sinh \left(\frac{\beta}{2}(\mathcal{E}_{u}^{-} - p_{0})  \right)\sinh \left(\frac{\beta}{2}(\mathcal{E}_{v}^{+} - p_{0})  \right)   }{\sinh \left(\frac{\beta}{2}\mathcal{E}_{u}^{-}  \right)\sinh \left(\frac{\beta}{2}\mathcal{E}_{v}^{+}\right) }  \right] \nonumber \\[0.5em]
& \quad\quad\quad\quad\quad\quad\quad -\theta \!\left( \mathcal{E}_{u}^{+} + \mathcal{E}_{v}^{+} - p_{0} \right)  \ln\!\left[ \frac{\sinh \left(\frac{\beta}{2}(\mathcal{E}_{u}^{+} - p_{0})  \right)\sinh \left(\frac{\beta}{2}(\mathcal{E}_{v}^{+} - p_{0})  \right)   }{\sinh \left(\frac{\beta}{2}\mathcal{E}_{u}^{+}  \right)\sinh \left(\frac{\beta}{2}\mathcal{E}_{v}^{+}\right) }  \right]   \nonumber \\[0.5em]
&\quad\quad\quad\quad\quad\quad\quad - \theta \!\left( \mathcal{E}_{u}^{-} + \mathcal{E}_{v}^{-} - p_{0} \right)  \left. \ln\!\left[ \frac{\sinh \left(\frac{\beta}{2}(\mathcal{E}_{u}^{-} - p_{0})  \right)\sinh \left(\frac{\beta}{2}(\mathcal{E}_{v}^{-} - p_{0})  \right)   }{\sinh \left(\frac{\beta}{2}\mathcal{E}_{u}^{-}  \right)\sinh \left(\frac{\beta}{2}\mathcal{E}_{v}^{-}\right) }  \right]   \right\}, 
\label{rho1}
\end{align} 
where the energy-dependent parameters $\mathcal{E}_{u}^{\pm}$ and $\mathcal{E}_{v}^{\pm}$ are defined:
\begin{align}
\mathcal{E}_{u}^{\pm} = \sqrt{(|\vec{q}|\pm|\vec{u}|)^{2}+s}, \quad\quad \mathcal{E}_{v}^{\pm} = \sqrt{(|\vec{q}|\pm|\vec{v}|)^{2}+t}.
\end{align} 
Performing an analogous calculation for the second delta component, one finds
\begin{align}
\rho_{\pi\pi}^{(2)}(p_{0}) &= \int_{0}^{\infty} \!\! ds \int_{0}^{\infty} \!\! dt \int_{0}^{\infty} d|\vec{q}|\,  \frac{|\vec{q}|^{4}}{24\pi^{5}\beta}   \int_{0}^{\infty} d|\vec{u}| \int_{0}^{\infty} d|\vec{v}| \ |\vec{u}||\vec{v}|\,\widetilde{D}_{\beta}(\vec{u},s) \, \widetilde{D}_{\beta}(\vec{v},t)  \nonumber \\[0.5em]
& \quad\quad\quad\quad\quad\quad  \Bigg\{ \theta \!\left(\mathcal{E}_{u}^{+} - \mathcal{E}_{v}^{-} - p_{0} \right)   \ln\!\left[ \frac{\sinh \left(\frac{\beta}{2}(\mathcal{E}_{u}^{+} - p_{0})  \right)\sinh \left(\frac{\beta}{2}(\mathcal{E}_{v}^{-} + p_{0})  \right)   }{\sinh \left(\frac{\beta}{2}\mathcal{E}_{u}^{+}  \right)\sinh \left(\frac{\beta}{2}\mathcal{E}_{v}^{-}\right) }  \right]   \nonumber \\[0.5em]
&  \quad\quad\quad\quad\quad\quad\quad + \theta \!\left( \mathcal{E}_{u}^{-} - \mathcal{E}_{v}^{+} - p_{0} \right)  \ln\!\left[ \frac{\sinh \left(\frac{\beta}{2}(\mathcal{E}_{u}^{-} - p_{0})  \right)\sinh \left(\frac{\beta}{2}(\mathcal{E}_{v}^{+} + p_{0})  \right)   }{\sinh \left(\frac{\beta}{2}\mathcal{E}_{u}^{-}  \right)\sinh \left(\frac{\beta}{2}\mathcal{E}_{v}^{+}\right) }  \right] \nonumber \\[0.5em]
& \quad\quad\quad\quad\quad\quad\quad -\theta \!\left( \mathcal{E}_{u}^{+} - \mathcal{E}_{v}^{+} - p_{0} \right)  \ln\!\left[ \frac{\sinh \left(\frac{\beta}{2}(\mathcal{E}_{u}^{+} - p_{0})  \right)\sinh \left(\frac{\beta}{2}(\mathcal{E}_{v}^{+} + p_{0})  \right)   }{\sinh \left(\frac{\beta}{2}\mathcal{E}_{u}^{+}  \right)\sinh \left(\frac{\beta}{2}\mathcal{E}_{v}^{+}\right) }  \right]   \nonumber \\[0.5em]
&\quad\quad\quad\quad\quad\quad\quad - \theta \!\left( \mathcal{E}_{u}^{-} - \mathcal{E}_{v}^{-} - p_{0} \right)  \left. \ln\!\left[ \frac{\sinh \left(\frac{\beta}{2}(\mathcal{E}_{u}^{-} - p_{0})  \right)\sinh \left(\frac{\beta}{2}(\mathcal{E}_{v}^{-} + p_{0})  \right)   }{\sinh \left(\frac{\beta}{2}\mathcal{E}_{u}^{-}  \right)\sinh \left(\frac{\beta}{2}\mathcal{E}_{v}^{-}\right) }  \right]   \right\}. 
\label{rho2}
\end{align} 
For the final contribution $\rho_{\pi\pi}^{(3)}(p_{0})$, one can see from Eq.~\eqref{conv} that this can in fact be related to $\rho_{\pi\pi}^{(1)}(p_{0})$ via the interchange $p_{0} \rightarrow -p_{0}$, and in particular:
\begin{align}
\rho_{\pi\pi}^{(3)}(p_{0}) = -\rho_{\pi\pi}^{(1)}(-p_{0}).
\label{relation1&3}
\end{align}
Moreover, due to Eqs.~\eqref{symmetry} and~\eqref{conv} it follows that $\rho_{\pi\pi}^{(2)}(p_{0})$ is anti-symmetric in $p_{0}$, which when combined with Eq.~\eqref{relation1&3} implies $\rho_{\pi\pi}(p_{0})=-\rho_{\pi\pi}(-p_{0})$, as expected. An important characteristic of this representation is that general support properties of the components $\rho_{\pi\pi}^{(i)}(p_{0})$ can be inferred from the delta terms appearing in Eq.~\eqref{conv}. In particular, since Eq.~\eqref{spec_rep} implies that $\widetilde{D}_{\beta}(\vec{u},s)$ is defined somewhere in the region $0 \leq s < \infty$, if follows that $\widetilde{D}_{\beta}(\vec{u},s)$ has support for $s \geq \nu$, where $\nu$ is some non-negative value. In the case of a massive theory, $\sqrt{\nu}$ is simply the mass gap $m$. By requiring that the arguments of the delta terms in Eq.~\eqref{conv} are non-vanishing, it therefore follows that $\rho_{\pi\pi}^{(1)}(p_{0})$, $\rho_{\pi\pi}^{(2)}(p_{0})$, and $\rho_{\pi\pi}^{(3)}(p_{0})$ are defined, respectively, in the following regions: 
\begin{align}
2\sqrt{\nu} \leq p_{0}  < \infty, \quad -\infty < p_{0} < \infty, \quad  -\infty < p_{0} < -2\sqrt{\nu}. \label{supp}
\end{align}
To our knowledge, these analytic properties of $\rho_{\pi\pi}(p_{0})$, in particular the representations in Eqs.~\eqref{rho1} and~\eqref{rho2}, are novel.   \\

\noindent
Now that we have explicit expressions for the components of $\rho_{\pi\pi}(p_{0})$, one can apply the Kubo relation in order to derive a spectral representation for $\eta_{0}$. In light of the support properties of $\rho_{\pi\pi}^{(i)}(p_{0})$ in Eq.~\eqref{supp}, if $\widetilde{D}_{\beta}(\vec{u},s)$ is defined such that $\nu>0$, then Eq.~\eqref{shear} implies that only the component $\rho_{\pi\pi}^{(2)}(p_{0})$ will provide a non-vanishing contribution to $\eta_{0}$. Under this assumption, after differentiating Eq.~\eqref{rho2}, shifting variables, and then taking the limit $p_{0}\rightarrow 0$ under the integral sign\footnote{The specific conditions under which the $p_{0}\rightarrow 0$ limit and the integrals can be swapped is outlined in the Appendix.}, the shear viscosity arising from the thermal asymptotic states takes the form
\begin{align}
\eta_{0} &=\int_{0}^{\infty} \!\! ds \int_{0}^{\infty} \!\! dt  \int_{0}^{\infty} d|\vec{u}| \int_{0}^{\infty} d|\vec{v}| \ \frac{|\vec{u}||\vec{v}|}{480\pi^{5}}\,\widetilde{D}_{\beta}(\vec{u},s) \, \widetilde{D}_{\beta}(\vec{v},t)  \int_{0}^{\infty} \!\! d|\vec{q}|  \, \frac{1}{e^{\beta\sqrt{|\vec{q}|^{2}+t}}-1}  \nonumber \\[0.5em]
&    \times\Bigg[  (|\vec{q}|+|\vec{v}|)^{4} \left\{ \epsilon  \!\left(\sqrt{s+(|\vec{q}|+|\vec{v}|+|\vec{u}|)^{2}}- \sqrt{t+|\vec{q}|^{2}}\right)  -\epsilon\! \left(\sqrt{s+(|\vec{q}|+|\vec{v}|-|\vec{u}|)^{2}}- \sqrt{t+|\vec{q}|^{2}} \right) \right\} \nonumber \\[0.5em]
&  -(|\vec{q}|-|\vec{v}|)^{4} \left\{ \epsilon \! \left(\sqrt{s+(|\vec{q}|-|\vec{v}|+|\vec{u}|)^{2}}- \sqrt{t+|\vec{q}|^{2}} \right)  -\epsilon\! \left(\sqrt{s+(|\vec{q}|-|\vec{v}|-|\vec{u}|)^{2}}- \sqrt{t+|\vec{q}|^{2}} \right) \right\} \Bigg].
\label{eta_1}
\end{align} 
In the $|\vec{q}|$ integrand there appears an explicit factor of the Bose-Einstein distribution 
\begin{align}
n(\omega_{\vec{q}}) = \frac{1}{e^{\beta\omega_{\vec{q}}}-1},
\label{BE}
\end{align}
with $\omega_{\vec{q}}=\sqrt{|\vec{q}|^{2}+t}$. One can make the parameter dependence of Eq.~\eqref{eta_1} manifest by rewriting this expression in terms of the following class of positive-valued integrals:
\begin{align}
&\mathcal{I}_{N}(R,a,b) =   \int_{0}^{b} \! d\hat{q} \ (\hat{q}-a)^{N} \, n\!\left(\sqrt{\hat{q}^{2}+R^{2}}\right),  \label{dimless_int} 
\end{align}
where the variables $\{R, a, b\}$ and $\hat{q}=|\vec{q}|/T$ are dimensionless. $\mathcal{I}_{N}(R,a,b)$ are proportional to the $|\vec{q}|$ moments of the Bose-Einstein distribution of a particle with mass $RT$ about the point $aT$, in the interval $[0,bT]$. After combining the various contributions from the sign functions in Eq.~\eqref{eta_1}, and applying the definition in Eq.~\eqref{dimless_int}, one ultimately finds
\begin{align}
\eta_{0} &=  \frac{T^{5}}{240\pi^{5}} \int_{0}^{\infty} \!\! ds \int_{0}^{\infty} \!\! dt\int_{0}^{\infty} d|\vec{u}| \int_{0}^{\infty} d|\vec{v}|  \, |\vec{u}||\vec{v}| \, \widetilde{D}_{\beta}(\vec{u},s) \, \widetilde{D}_{\beta}(\vec{v},t) \nonumber \\[0.5em]
& \quad\times \Bigg[  4\left[1+\epsilon(|\vec{u}|-|\vec{v}|)  \right]\left\{\frac{|\vec{v}|}{T} \, \mathcal{I}_{3}\!\left( \! \frac{\sqrt{t}}{T}, \, 0, \infty  \! \right)  +  \frac{|\vec{v}|^{3}}{T^{3}} \, \mathcal{I}_{1}\!\left( \! \frac{\sqrt{t}}{T}, \, 0, \infty \!\right) \right\}   \nonumber \\[0.5em]
& \quad\quad +   \left\{  \mathcal{I}_{4}\!\left( \! \frac{\sqrt{t}}{T},\frac{|\vec{v}|}{T},\frac{s-t +(|\vec{u}|+|\vec{v}|)^{2}}{2(|\vec{u}|+|\vec{v}|)T}\right) + \epsilon(|\vec{u}|-|\vec{v}|)\, \mathcal{I}_{4}\!\left( \! \frac{\sqrt{t}}{T},\frac{|\vec{v}|}{T},\frac{s-t+(|\vec{v}|-|\vec{u}|)^{2}}{2(|\vec{v}|-|\vec{u}|)T}\right) \right\} \Bigg].
\label{shear_general_st}  
\end{align}
Equation~\eqref{shear_general_st} explicitly demonstrates that the model dependence of $\eta_{0}$ factorises, and is completely determined by the form of the thermal spectral density $\widetilde{D}_{\beta}$. Since the integral kernel function in square brackets is fixed, one can use the properties of this function to establish general constraints on $\eta_{0}$. In particular, in the Appendix we prove the following result:
\begin{align}
\textit{If the KMS condition holds} \quad \Longrightarrow \quad \textit{$\eta_{0}$ is finite.}
\label{cond}   
\end{align}  
Therefore, if a system exists in a state of thermal equilibrium, this is sufficient to guarantee that $\eta_{0}$ is a meaningful observable. In the next section we will use Eq.~\eqref{shear_general_st} in order to calculate the explicit form of $\eta_{0}$ in specific examples, and discuss the implications of the condition in Eq.~\eqref{cond}.

\section{Shear viscosity of thermal scalar particles}
\label{shear_part}

\subsection{Spectral decomposition}
\label{spec_decomp}

The analysis in Secs.~\ref{sec:outline} and~\ref{shear_sec} demonstrates that the thermal spectral density plays an essential role in governing the dynamics of local QFTs at finite temperature, and in particular the behaviour of the shear viscosity. For non-vanishing temperatures, it is expected that the singular structure of $\widetilde{D}_{\beta}(\vec{u},s)$ in the variable $s$ is preserved relative to the vacuum theory, and hence the discrete and continuous contributions can be separated~\cite{Bros:1992ey}. In particular, this means that if a theory has a known particle state of mass $m$ at zero temperature, then $\widetilde{D}_{\beta}(\vec{u},s)$ has the following decomposition~\cite{Bros:2001zs}:
\begin{align}
\widetilde{D}_{\beta}(\vec{u},s)= \widetilde{D}_{m,\beta}(\vec{u})\, \delta(s-m^{2}) + \widetilde{D}_{c, \beta}(\vec{u},s),
\label{decomp}
\end{align} 
where $\widetilde{D}_{c, \beta}(\vec{u},s)$ is continuous in the variable $s$. This decomposition provides a natural description for the properties of particles in a thermal medium since Eqs.~\eqref{W_rep} and~\eqref{spec_rep} imply that the $\vec{u}$-dependent coefficient $\widetilde{D}_{m,\beta}(\vec{u})$ causes the correlation functions to have contributions outside of the mass shell $p^{2}=m^{2}$. In this sense, the rest mass $m$ of the particle state is screened, and $\widetilde{D}_{m,\beta}(\vec{u})$ has the interpretation of a thermal damping factor, the behaviour of which is fixed by the underlying dynamics between the particle and the constituents in the thermal background state. \\

\noindent
In Ref.~\cite{Bros:2001zs}, an important finding was made with regard to damping factors and their connection to the properties of asymptotic states. The authors proved that if the thermal spectral density of a real scalar field $\phi(x)$ satisfies Eq.~\eqref{decomp}, then the discrete particle component will dominate the behaviour of correlation functions in the asymptotic temporal limit. Moreover, as discussed in Sec.~\ref{shear_sec}, in this limit they demonstrated that all correlation functions are expressible in terms of sums of products of two-point functions, and in this sense the asymptotic contributions have the structure of quasi-free states. By introducing an asymptotic scalar field $\phi_{0}(x)$ satisfying a modified commutator algebra, the authors showed that the universal structure of these quasi-free states can be captured by these fields in a model-independent manner. By further demanding that $\phi_{0}(x)$ satisfies a specific asymptotic field equation, in particular requiring that the operator
\begin{align}
(\partial^{2}+m^{2})\phi_{0}(x) + \sum_{k=2}^{K}g_{k}\phi_{0}^{k}(x)
\label{asymptotic_eq}
\end{align}
is suppressed in all correlation functions in the asymptotic limit $x_{0}\rightarrow \pm \infty$ for some choice of (temperature-independent) parameters $m$ and $g_{k}$, this leads to the remarkable result that the thermal two-point function of $\phi_{0}(x)$ is \textit{uniquely} determined. Since this two-point function coincides with that of the full interacting field $\phi(x)$ for $x_{0}\rightarrow \pm \infty$, and the latter is dominated by the damping factor in this limit, this procedure therefore fixes the form of $\widetilde{D}_{m,\beta}(\vec{u})$ in terms of the parameters $m$ and $g_{k}$, which correspond to the mass and coupling strength experienced by the asymptotic states. Physically this makes sense since these parameters represent genuine observables, and hence one would expect the full non-perturbative correlation functions to be parametrised by them, as one indeed finds in phenomenological approaches such as the QCD sum rules, where the spectral density is fixed in terms of the masses and decay constants of the (asymptotic) hadronic states~\cite{Shifman:1978by}.  \\

\noindent
Since the discrete particle component in Eq.~\eqref{decomp} dominates the behaviour of the two-point function in the asymptotic temporal limit, the shear viscosity associated with the thermal particle states can therefore be calculated by making the substitution $\widetilde{D}_{\beta}(\vec{u},s) \rightarrow \widetilde{D}_{m,\beta}(\vec{u})\, \delta(s-m^{2})$ in Eq.~\eqref{shear_general_st}. As outlined in this section, despite their non-perturbative nature, the structure of the damping factors $\widetilde{D}_{m,\beta}$ is actually fixed by the dynamics of the asymptotic states, and hence one can use Eq.~\eqref{shear_general_st} in order to derive an explicit expression for $\eta_{0}$. In the remainder of this section we will explore the characteristics of $\eta_{0}$ in different models.

\subsection{Free scalar theory}
\label{free}

For free theories it turns out that the structure of the thermal commutator is completely independent of the thermal state $|\Omega_{\beta}\rangle$, and hence coincides with the zero temperature expression. In position space the damping factor therefore has the form $D_{m,\beta}(\vec{x})=1$, and hence
\begin{align}
\widetilde{D}_{m,\beta}(\vec{u}) = (2\pi)^{3}\delta^{3}(\vec{u}).
\label{D_free}
\end{align}
Applying Eqs.~\eqref{rho1} and~\eqref{rho2}, one finds 
\begin{align}
\rho_{\pi\pi}(p_{0}) = \frac{\left[p_{0}^{2} - 4m^{2} \right]^{3}}{96 \pi p_{0}^{2}}  \coth\left(\frac{\beta p_{0}}{4}\right)\left[ \theta(p_{0}-2m) + \theta(-p_{0}-2m)   \right],
\label{rhopipi_free}
\end{align}
where, in particular, $\rho_{\pi\pi}^{(2)}(p_{0})=0$. Since Eq.~\eqref{rhopipi_free} is only non-vanishing for $p_{0}\geq 2m$, it immediately follows from Eq.~\eqref{shear} that the shear viscosity of a free field is zero, as pointed out in Ref.~\cite{Jeon:1992kk}.

\subsection{Thermal particle shear viscosity in $\phi^{4}$ theory}

At zero temperature, the negative and positive coupling regimes of $\phi^{4}$ theory represent two very different phases of the theory. The former is unstable but known to be asymptotically free~\cite{Symanzik:1973hx}, whereas the latter has a ground state but is expected to be trivial in 3+1 dimensions~\cite{Glimm:1987ng}. At finite temperature, less is known about the non-perturbative characteristics of the theory, although progress has been made in understanding the spectral structure of the two-point function $\langle \Omega_{\beta}|\phi(x)\phi(y)|\Omega_{\beta}\rangle$. In particular, by applying the procedure outlined in Sec.~\ref{spec_decomp}, and demanding that the equation of motion\footnote{In Ref.~\cite{Bros:2001zs} the $\frac{1}{3!}$ factor was not included in Eq.~\eqref{asymptotic_eq_phi4}, but here and throughout this paper we will use this in order to be consistent with the perturbative convention.} operator
\begin{align}
(\partial^{2}+m^{2})\phi_{0}(x) + \frac{\lambda}{3!}  \phi_{0}^{3}(x)
\label{asymptotic_eq_phi4}
\end{align}
is suppressed in all correlation functions of the asymptotic field $\phi_{0}$, in Ref.~\cite{Bros:2001zs} the authors were able to explicitly calculate the form of the damping factor $\widetilde{D}_{m,\beta}(\vec{u})$ for both $\lambda<0$ and $\lambda>0$. As emphasised in Sec.~\ref{spec_decomp}, $m$ and $\lambda$ represent physical parameters in the theory: the zero-temperature mass, and the coupling of the asymptotic states.

\subsubsection{Negative coupling shear viscosity}
\label{Phi4_neg}

Setting $\lambda<0$, the condition that the field operator in Eq.~\eqref{asymptotic_eq_phi4} vanishes asymptotically implies that the position space damping factor has the form~\cite{Bros:2001zs}     
\begin{align}
D_{m,\beta}(\vec{x}) = \frac{\sin(\kappa |\vec{x}|)}{\kappa |\vec{x}|}, 
\label{D_phi4_x}
\end{align}
where $\kappa$ is a positive function of the parameters $\left\{\lambda,\beta,m\right\}$ and is related in the following manner to the Lorentz invariant integral of the Bose-Einstein distribution $n(E_{\vec{q}})$ of a particle with energy $E_{\vec{q}}= \sqrt{|\vec{q}|^{2}+m^{2}}$:
\begin{align}
\kappa = \sqrt{|\lambda|} \, \sqrt{  \int \!\! \frac{d^{3}\vec{q}}{(2\pi)^{3}2E_{\vec{q}}} n(E_{\vec{q}}) }.
\label{kappa}
\end{align}
In the region $|\vec{x}|\ll \kappa^{-1}$, one can see from Eq.~\eqref{D_phi4_x} that the damping factor approaches the free field expression, namely $D_{m,\beta}(\vec{x}) \rightarrow 1$. Since $|\vec{x}|$ is the spatial separation of the fields, $\ell= \kappa^{-1}$ can therefore be interpreted as the mean free path of the (asymptotic) particles in the thermal state~\cite{Bros:2001zs}. This matches physical expectations, since $\ell\rightarrow \infty$ in the weak coupling ($|\lambda| \rightarrow 0$) and small temperature ($\beta\rightarrow \infty$) limits. In the special case $m=0$, one has the exact expression: $\ell = 2\sqrt{6}|\lambda|^{-\frac{1}{2}}\beta$. Taking the Fourier transform of Eq.~\eqref{D_phi4_x} gives
\begin{align}
\widetilde{D}_{m,\beta}(\vec{u}) = \frac{2\pi^{2}}{\kappa^{2}} \left[  \delta(|\vec{u}|-\kappa) + \delta(|\vec{u}|+\kappa)   \right].
\label{D_phi4}  
\end{align}
Equation~\eqref{D_phi4} reduces to the free theory result of Eq.~\eqref{D_free} as the coupling approaches zero, which is what one would expect if the theory were asymptotically free, like its zero temperature limit~\cite{Symanzik:1973hx}.  \\

\begin{figure}[H]
\centering
\includegraphics[width=0.8\columnwidth]{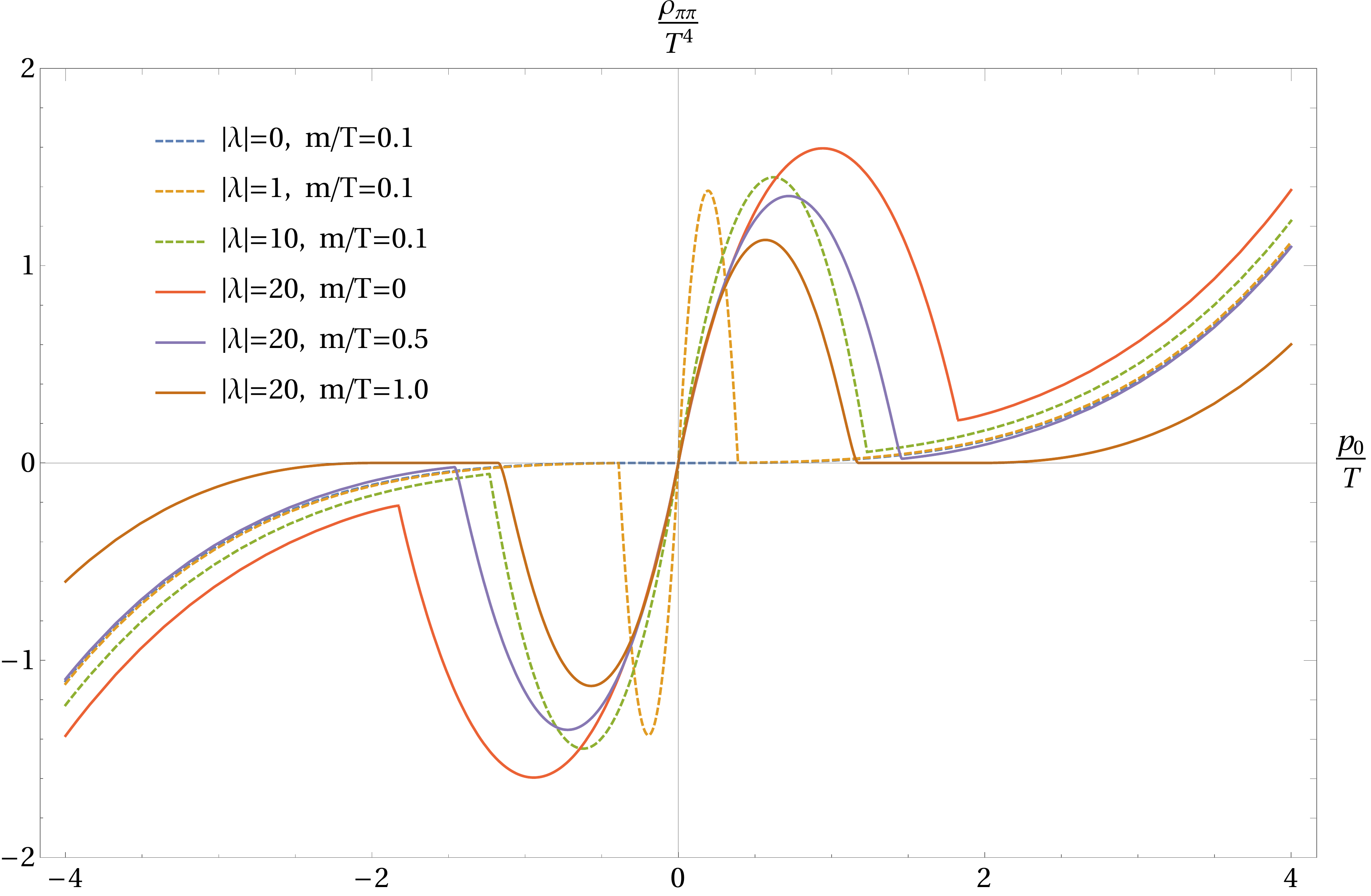}
\caption{$\rho_{\pi\pi}(p_{0})/T^{4}$ plotted as a function of $p_{0}/T$ for fixed $m/T$ (dashed lines) and fixed $|\lambda|$ (solid lines).}
\label{fig_combined}
\end{figure}

\noindent
Now that we have the explicit form of the damping factor, one can apply Eqs.~\eqref{rho1} and~\eqref{rho2} in order to calculate the thermal particle contribution to $\rho_{\pi\pi}(p_{0})$. In the case of the first spectral component, one finds 
\begin{align}
&\rho_{\pi\pi}^{(1)}(p_{0}) = \frac{\theta(p_{0}-2\sqrt{\kappa^{2}+m^{2}})}{6\pi\beta\kappa^{2}}   \int_{\frac{1}{2}\alpha}^{\kappa+\frac{1}{2}\gamma}  d|\vec{q}|\  |\vec{q}|^{4}   \ln\!\left[ \frac{\sinh \left(\frac{\beta}{2}(\sqrt{(|\vec{q}|-\kappa)^{2}+m^{2}}  - p_{0})  \right) }{\sinh \left(\frac{\beta}{2}(\sqrt{(|\vec{q}|-\kappa)^{2}+m^{2}})  \right) }  \right]    \nonumber \\[0.5em]
& \quad\quad\quad\quad +\frac{\theta(p_{0}-2m)\theta(2\sqrt{\kappa^{2}+m^{2}}- p_{0})}{6\pi\beta\kappa^{2}}  \int_{\kappa-\frac{1}{2}\gamma}^{\kappa+\frac{1}{2}\gamma} d|\vec{q}|\  |\vec{q}|^{4}   \ln\!\left[ \frac{\sinh \left(\frac{\beta}{2}(\sqrt{(|\vec{q}|-\kappa)^{2}+m^{2}}  - p_{0})  \right) }{\sinh \left(\frac{\beta}{2}(\sqrt{(|\vec{q}|-\kappa)^{2}+m^{2}})  \right) }  \right] \nonumber \\[0.5em]
& \quad\quad\quad\quad - \frac{\theta(p_{0}-2\sqrt{\kappa^{2}+m^{2}})}{6\pi\beta\kappa^{2}} \int_{\frac{1}{2}\gamma -\kappa}^{\frac{1}{2}\alpha} d|\vec{q}|\  |\vec{q}|^{4}  \ln\!\left[ \frac{\sinh \left(\frac{\beta}{2}(\sqrt{(|\vec{q}|+\kappa)^{2}+m^{2}}  - p_{0})  \right) }{\sinh \left(\frac{\beta}{2}(\sqrt{(|\vec{q}|+\kappa)^{2}+m^{2}})  \right) }  \right],  
\label{rho1_phi4}
\end{align} 
where the threshold parameters $\alpha$ and $\gamma$ are defined
\begin{align}
\alpha =\sqrt{\frac{4p_{0}^{2}(\kappa^{2}+m^{2})-p_{0}^{4}}{4\kappa^{2}-p_{0}^{2}}}, \quad\quad \gamma=\sqrt{p_{0}^{2}-4m^{2}}.
\end{align} 
One can immediately see from Eq.~\eqref{rho1_phi4} that $\rho_{\pi\pi}^{(1)}(p_{0})$ has support for $p_{0} \geq 2m$, which agrees with the general result in Eq.~\eqref{supp}. For the second spectral component:
\begin{align}
\rho_{\pi\pi}^{(2)}(p_{0}) &= \frac{\theta(p_{0})\theta(2\kappa-p_{0})}{6\pi\beta\kappa^{2}}   \int_{\frac{1}{2}\alpha}^{\infty} d|\vec{q}|\  |\vec{q}|^{4}  \left\{  \ln\!\left[ \frac{\sinh \left(\frac{\beta}{2}(\sqrt{(|\vec{q}|-\kappa)^{2}+m^{2}}  + p_{0})  \right)}{\sinh \left(\frac{\beta}{2}(\sqrt{(|\vec{q}|-\kappa)^{2}+m^{2}})  \right)}  \right] \right. \nonumber \\[0.5em]  
&\quad\quad\quad\quad\quad\quad\quad\quad\quad\quad\quad\quad\quad\quad\quad\quad + \left.\ln\!\left[ \frac{\sinh \left(\frac{\beta}{2}(\sqrt{(|\vec{q}|+\kappa)^{2}+m^{2}}  - p_{0})  \right) }{\sinh \left(\frac{\beta}{2}(\sqrt{(|\vec{q}|+\kappa)^{2}+m^{2}})  \right) }  \right]  \right\}       \nonumber \\[0.5em]
&\quad - \frac{\theta(-p_{0})\theta(2\kappa+p_{0})}{6\pi\beta\kappa^{2}}    \int_{\frac{1}{2}\alpha}^{\infty} d|\vec{q}|\  |\vec{q}|^{4}  \left\{  \ln\!\left[ \frac{\sinh \left(\frac{\beta}{2}(\sqrt{(|\vec{q}|-\kappa)^{2}+m^{2}}  - p_{0})  \right)}{\sinh \left(\frac{\beta}{2}(\sqrt{(|\vec{q}|-\kappa)^{2}+m^{2}})  \right)}  \right] \right. \nonumber \\[0.5em]  
&\quad\quad\quad\quad\quad\quad\quad\quad\quad\quad\quad\quad\quad\quad\quad\quad + \left.\ln\!\left[ \frac{\sinh \left(\frac{\beta}{2}(\sqrt{(|\vec{q}|+\kappa)^{2}+m^{2}}  + p_{0})  \right) }{\sinh \left(\frac{\beta}{2}(\sqrt{(|\vec{q}|+\kappa)^{2}+m^{2}})  \right) }  \right]  \right\},  
\label{rho2_phi4}
\end{align} 
and hence $\rho_{\pi\pi}^{(2)}(p_{0})$ is non-vanishing in the region $[-2\kappa,2\kappa]$, which is again consistent with Eq.~\eqref{supp}. The $|\vec{q}|$ integrals in $\rho_{\pi\pi}^{(1)}(p_{0})$ and $\rho_{\pi\pi}^{(2)}(p_{0})$ are convergent for all values of the asymptotic parameters, and when $m=0$ they can in fact be evaluated in terms of elementary functions. Due to the general anti-symmetry relation in Eq.~\eqref{relation1&3}, the full spectral function therefore has the form
\begin{align}
\rho_{\pi\pi}(p_{0})=\rho_{\pi\pi}^{(1)}(p_{0})+\rho_{\pi\pi}^{(2)}(p_{0})-\rho_{\pi\pi}^{(1)}(-p_{0}).
\end{align}
\ \\
\noindent
In Fig.~\ref{fig_combined}, the dimensionless normalised spectral function $\rho_{\pi\pi}(p_{0})/T^{4}$ is plotted as a function of $p_{0}/T$ for both fixed $m/T$ and coupling strength $|\lambda|$. For $\lambda=0$, the plot coincides with the exact free field expression in Eq.~\eqref{rhopipi_free}. When $m/T>0$ and $|\lambda|>0$, the $|\vec{q}|$ integrals in Eqs.~\eqref{rho1_phi4} and~\eqref{rho2_phi4} are evaluated numerically, and in the special case $m/T=0$ the integrals are solvable in terms of elementary functions. One can see from Fig.~\ref{fig_combined} that the presence of non-trivial interactions causes the appearance of resonant peaks at $p_{0}=\pm\kappa$. Physically, this corresponds to the point at which the energy of the particles approaches the inverse of their mean free path $\ell^{-1}$. As $m/T$ decreases, the resonance peak grows, and $\rho_{\pi\pi}(p_{0})/T^{4}$ moves off the $p_{0}/T$ axis, becoming non-zero for all $p_{0}/T \neq 0$. This occurs in the region $m \leq \kappa$, where the particle rest mass is smaller than the interaction energy with the thermal background.

\ \\
\noindent
Applying the spectral representation in Eq.~\eqref{shear_general_st}, one can now use the form of the damping factor in Eq.~\eqref{D_phi4} to calculate the thermal particle shear viscosity $\eta_{0}$. In order to make the dimensional dependence of this representation explicit, we rewrite the expression in Eq.~\eqref{kappa} for $\kappa$ in terms of the dimensionless parameter $r=m/T$
\begin{align}
\kappa = T\sqrt{|\lambda|} K(r), \quad\quad  K(r) =  \sqrt{  \int \!\! \frac{d^{3}\hat{\vec{q}}}{(2\pi)^{3}2\sqrt{|\hat{\vec{q}}|^{2}+r^{2}}} \, \frac{1}{e^{\sqrt{|\hat{\vec{q}}|^{2}+r^{2}}}-1}  },
\label{K_def}
\end{align} 
and also define the following rescaled version of the functions in Eq.~\eqref{dimless_int}: 
\begin{align}
\mathcal{K}_{N}(r,a,b) = \left[K(r)\right]^{2-N}\mathcal{I}_{N}(r,a,b). 
\label{Kint}   
\end{align}
Finally, after setting $\widetilde{D}_{\beta}(\vec{u},s) = \widetilde{D}_{m,\beta}(\vec{u})\, \delta(s-m^{2})$ in Eq.~\eqref{shear_general_st}, and using Eqs.~\eqref{K_def} and~\eqref{Kint}, the thermal particle shear viscosity takes the form
\begin{align}
\eta_{0} &=  \frac{T^{3}}{15\pi}  \! \left[  \frac{ \mathcal{K}_{3}\!\left(  \frac{m}{T}, \, 0, \infty  \right)}{\sqrt{|\lambda|}}  +  \sqrt{|\lambda|} \, \mathcal{K}_{1}\!\left(\frac{m}{T},  \, 0, \infty \right)    + \frac{\mathcal{K}_{4}\!\left(  \frac{m}{T},\sqrt{|\lambda|}K\!\left(\frac{m}{T}\right),\sqrt{|\lambda|}K\!\left(\frac{m}{T}\right) \right)}{4|\lambda|}   \right].
\label{shear_3}  
\end{align}
\noindent
Equation~\eqref{shear_3} implies that for fixed coupling strength $|\lambda|$ the temperature dependence of $\eta_{0}/T^{3}$ is entirely controlled by functions of $m/T$, similarly to what one finds in perturbative calculations~\cite{Jeon:1994if,Jeon:1995zm}. However, in contrast to the perturbative case, Eq.~\eqref{shear_3} is valid for \textit{all} values of $|\lambda|$. \\

\begin{figure}[H]
\centering
\includegraphics[width=0.65\columnwidth]{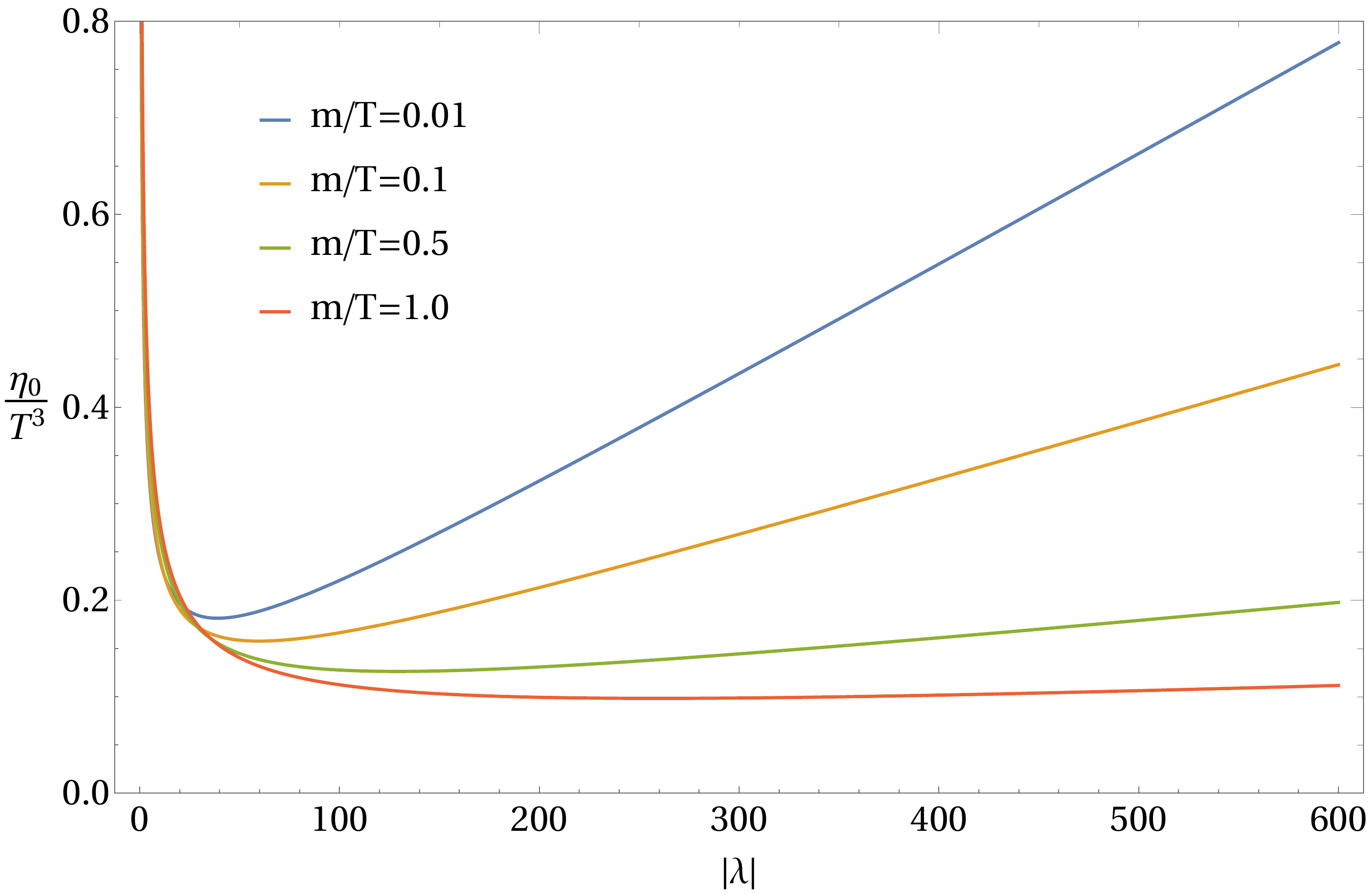}
\caption{$\eta_{0}/T^{3}$ plotted as a function of $|\lambda|$ for varying $m/T$.}
\label{fig2}
\end{figure}

\noindent
In Fig.~\ref{fig2}, we plot $\eta_{0}/T^{3}$ as a function of $|\lambda|$ for different values of $m/T$. One can see that $\eta_{0}/T^{3}$ diverges for both small and large values of $|\lambda|$, and at some finite value of $\lambda$ there exists a global minimum. In particular, using Eq.~\eqref{shear_3} together with the boundedness properties of the functions $\mathcal{K}_{N}$, one finds that $\eta_{0}$ has the following asymptotic behaviour:
\begin{align} 
&\eta_{0} \, \sim \, \frac{T^{3}}{15\pi\sqrt{|\lambda|}} \, \mathcal{K}_{3}\!\left(  \frac{m}{T}, \, 0, \infty  \right),  \hspace{10mm} |\lambda|\rightarrow 0, \label{shear_small}  \\
&\eta_{0} \, \sim \,\frac{|\lambda| T^{3}}{60\pi} \, \mathcal{K}_{0}\!\left(  \frac{m}{T}, \, 0, \infty  \right),  \hspace{16mm} |\lambda|\rightarrow \infty,
\label{shear_large}  
\end{align}
where $\sim$ refers to the leading terms in the small and large coupling asymptotic expansions.
So far we have implicitly assumed that $m>0$, since this is required in order to apply the spectral representation in Eq.~\eqref{shear_general_st}. In fact, from the analysis in the Appendix it turns out that the condition $m>0$ is actually both necessary and sufficient for $\eta_{0}$ to be finite. Therefore, even though $\rho_{\pi\pi}(p_{0})$ is well-defined when $m=0$, $\eta_{0}$ is divergent. This divergence is reflected in the large coupling asymptotic behaviour in Eq.~\eqref{shear_large}, where one finds 
\begin{align}
\mathcal{K}_{0}\!\left(  \frac{m}{T}, \, 0, \infty  \right)   \sim \,  -\frac{1}{24} \ln \left(\frac{m}{2T}  \right), \quad\quad\quad m \rightarrow 0.
\end{align}
In this sense $m$ acts as an infrared regulator, guaranteeing the finiteness of $\eta_{0}$ for all non-vanishing temperatures and coupling strengths. \\

\noindent
Equation~\eqref{shear_small} demonstrates that for fixed temperatures and masses, $\eta_{0}$ diverges in the zero coupling limit. On initial inspection this appears to contradict the conclusions of Sec.~\ref{free}, where one finds for a free theory that $\eta=0$. However, as we will now demonstrate, the expectation that the shear viscosity of an interacting theory is a continuous perturbation of the free theory turns out to be false. This can be seen from the explicit structure of $\rho_{\pi\pi}^{(1)}(p_{0})$ and $\rho_{\pi\pi}^{(2)}(p_{0})$ in Eqs.~\eqref{rho1_phi4} and~\eqref{rho2_phi4}, in particular their support properties. As previously noted in Sec.~\ref{shear_sec}, for $m>0$ the first spectral component $\rho_{\pi\pi}^{(1)}(p_{0})$ does not contribute to the shear viscosity since it only has support for $p_{0} \geq 2m$. For the second spectral function component, one can see that $\rho_{\pi\pi}^{(2)}(p_{0})$ has an overall $\theta(p_{0})\theta(2\kappa -p_{0})$ coefficient for positive $p_{0}$. This coefficient remains for $\frac{d \rho_{\pi\pi}^{(2)}}{d p_{0}}$ because its $p_{0}$ derivative gives no contribution. Due to Eq.~\eqref{K_def}, $\lambda=0$ implies $\kappa=0$, and hence in the free theory this coefficient exactly vanishes. In the interacting theory though, $\kappa>0$ for $T>0$, and so taking the $p_{0} \rightarrow 0$ limit of $\frac{d \rho_{\pi\pi}^{(2)}}{d p_{0}}$ leads to a non-vanishing contribution to $\eta_{0}$. This implies that the free theory shear viscosity cannot be recovered from the zero coupling ($\kappa \rightarrow 0$) limit of the interacting theory result because the limits $p_{0}\rightarrow 0$ and $|\lambda| \rightarrow 0$ are non-commutative, due to the appearance of the coefficient $\theta(p_{0})\theta(2\kappa -p_{0})$. 
\begin{figure}[H]
\centering
\includegraphics[width=0.9\columnwidth]{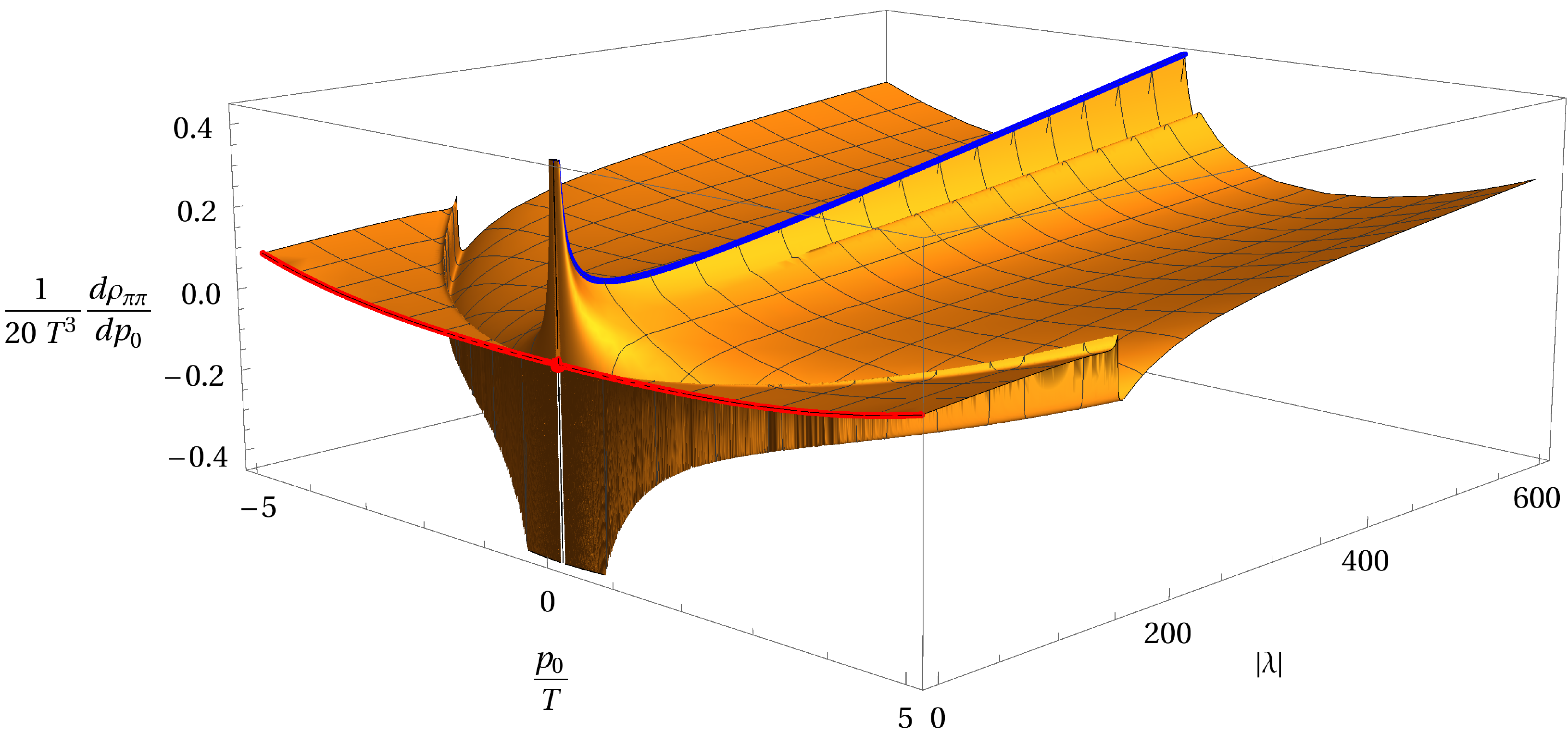}
\caption{$\frac{1}{20T^{3}}\!\frac{d \rho_{\pi\pi}}{d p_{0}}$ plotted as a function of $p_{0}/T$ and $|\lambda|$ for $m/T=0.1$. The blue and red lines correspond to the fixed surfaces $\frac{p_{0}}{T}=0$ and $\lambda=0$, respectively.}
\label{fig4}
\end{figure} 
\noindent
In Fig.~\ref{fig4} $\frac{1}{20T^{3}}\!\frac{d \rho_{\pi\pi}}{d p_{0}}$ is plotted as a function of $p_{0}/T$ and $|\lambda|$ for $m/T=0.1$. The blue line is the fixed surface $p_{0}/T=0$, corresponding to $\eta_{0}/T^{3}$, and the red line is the free field expression. The non-commutativity of the zero-energy and zero-coupling limits is reflected in the fact that the value of the function at $(p_{0}/T,|\lambda|)=(0,0)$ depends upon the direction in which the limit is taken, and so $(0,0)$ represents a non-analytic point. Keeping $\lambda=0$ fixed, and then $p_{0}/T \rightarrow 0$, corresponds to travelling along the red line towards the red point, but setting $p_{0}/T = 0$ first, and then taking $|\lambda| \rightarrow 0$, amounts to moving along the blue line towards the $p_{0}/T$ axis. In the first case, the function $\frac{1}{20T^{3}}\frac{d \rho_{\pi\pi}}{d p_{0}}$ vanishes, and hence $\eta_{0}=0$ for a free theory, whereas in the second case the function diverges, which implies $\lim_{|\lambda| \rightarrow 0}\eta_{0} = \infty$.

\subsubsection{Positive coupling shear viscosity}
\label{Phi4_pos}

Setting $\lambda>0$, and following the same procedure as in the negative coupling case, the position space damping factor takes the form~\cite{Bros:2001zs}    
\begin{align}
D_{m,\beta}(\vec{x}) = \frac{e^{-\kappa |\vec{x}|}}{\kappa_{0} |\vec{x}|},  
\label{D_phi4_x2}
\end{align}
where $\kappa$ is the same expression as in Eq.~\eqref{kappa}. Unlike the $\lambda<0$ case, Eq.~\eqref{D_phi4_x2} involves an additional fixed energy scale $\kappa_{0}$. This arises because the damping factor solutions have a more singular structure due to the divergent ultraviolet properties of the quartic interaction. The asymptotic fields must therefore be renormalised, which results in the introduction of the scale $\kappa_{0}$. Taking the Fourier transform of Eq.~\eqref{D_phi4_x2} gives
\begin{align}
\widetilde{D}_{m,\beta}(\vec{u}) = \frac{4\pi}{\kappa_{0}\left(|\vec{u}|^{2}+\kappa^{2} \right)}. 
\label{D_phi4_2}  
\end{align}
One can immediately see that unlike the $\lambda<0$ damping factor, this expression does not approach the free field result of Eq.~\eqref{D_free} in the zero coupling limit. Moreover, by setting $\widetilde{D}_{m,\beta}(\vec{u},s)=\widetilde{D}_{m,\beta}(\vec{u})\delta(s-m^{2})$, one finds that
\begin{align}
\textit{$\eta_{0}$ is divergent for $\lambda>0$.}
\label{div_cond}  
\end{align}
Since $\widetilde{D}_{m,\beta}(\vec{u},s)$ satisfies all of the assumptions leading to the general condition in Eq.~\eqref{cond}, it therefore follows from the contrapositive of this condition that the KMS condition must be violated. This is indeed the case, as outlined in the Appendix. Physically, this implies that the positive coupling quartic interaction is not consistent with the existence of thermal equilibrium. Whilst it might appear surprising that $\eta_{0}$ is divergent, since perturbative calculations have been performed in this model~\cite{Jeon:1994if,Jeon:1995zm}, it should be noted that the divergence of $\eta_{0}$ is a non-perturbative result, and so does not necessarily need to coincide with perturbative expectations. This is certainly the case at $T=0$, where the renormalised perturbative series is seemingly disconnected from the full (trivial) non-perturbative solution~\cite{Strocchi:2013awa}. In the strong coupling calculations that have been performed with $\lambda>0$, for example using variational methods~\cite{Aarts:2003bk,Aarts:2004sd}, an ultraviolet cutoff is assumed. In light of the condition in Eq.~\eqref{div_cond} one would therefore expect that the shear viscosity must ultimately diverge in these calculations in the limit of cutoff removal.

\section{Conclusions} 
\label{concl}

Local formulations of QFT at finite temperature imply the existence of non-perturbative constraints on the structure of thermal correlation functions. In this work, we use these constraints in order to derive a spectral representation for the shear viscosity arising from scalar thermal asymptotic states, $\eta_{0}$. Using this representation, we calculate the explicit form of $\eta_{0}$ in $\phi^{4}$ theory at both positive and negative coupling. For negative coupling, we find that $\eta_{0}$ possesses a global minimum, and grows unbounded for both small and large values of the coupling strength, whereas for positive coupling $\eta_{0}$ diverges for all parameter values. We subsequently demonstrate that the divergence of $\eta_{0}$ in the positive coupling theory is a reflection of the fact that the quartic interaction is not consistent with the existence of thermal equilibrium. \\

\noindent
Since the constraints of local QFT also apply to theories with non-scalar fields, as well as those with a non-vanishing chemical potential, this work represents a first step in understanding the analytic structure of in-medium observables for more complex theories of physical interest, including gauge theories like QCD. Although this work has focussed solely on the non-perturbative characteristics of thermal correlation functions in Minkowski spacetime, it turns out that the corresponding Euclidean quantities also possess specific constraints. By utilising these constraints in conjunction with non-perturbative results from lattice QFT~\cite{Nakamura:2004sy,Meyer:2007ic,Meyer:2011gj,Astrakhantsev:2017nrs}, or functional methods~\cite{Tripolt:2013jra,Tripolt:2014wra,Pawlowski:2015mia,Pawlowski:2017gxj,Tripolt:2020irx,Horak:2020eng}, this could provide new insights into both the spectral and transport properties of local QFTs. This will be the subject of forthcoming work.

\section*{Acknowledgements}

The authors would like to thank Detlev Buchholz, Guy Moore, Jan Horak, and Nicolas Wink for useful discussions and input. The work of P.~L. and D.~H.~R. is supported by the Deutsche Forschungsgemeinschaft (DFG, German Research Foundation) through the Collaborative Research Center CRC-TR 211 ``Strong-interaction matter under extreme conditions'' -- Project No. 315477589-TRR 211. R.-A.~T. is supported by the Austrian Science Fund (FWF) through Lise Meitner Grant No. M 2908-N. The work of J.~M.~P is supported by the DFG under Germany's Excellence Strategy Grant No. EXC 2181/1 - 390900948 (the Heidelberg STRUCTURES Excellence Cluster), and under the Collaborative Research Centre SFB 1225 (ISOQUANT), and the BMBF Grant No. 05P18VHFCA.

\appendix

\section{Boundedness of the shear viscosity}
\label{bound}

Here we will prove the result in Eq.~\eqref{cond}. First, it follows from the definition in Eq.~\eqref{dimless_int} that the functions $\mathcal{I}_{N}(R,a,b)$ are continuous in their arguments for positive $N$, and vanish in the limit $R\rightarrow \infty$. In the particular case of the functions $\mathcal{I}_{1}(R,0,\infty)$, $\mathcal{I}_{3}(R,0,\infty)$, and $\mathcal{I}_{4}(R,a,b)$, appearing in Eq.~\eqref{shear_general_st}, the first two are finite for $R\rightarrow 0$, and the last one diverges in this limit. However, since we assume that $\widetilde{D}_{\beta}(\vec{u},s)$ is defined for $s\in [\nu,\infty)$, where $\nu>0$, the singularity of $\mathcal{I}_{4}(R,a,b)$ is not contained within the integration range. Taken together, these properties imply that the coefficient in square brackets in Eq.~\eqref{shear_general_st} is bounded from above by polynomials in $|\vec{u}|$, $|\vec{v}|$, $s$, and $t$, and approaches a constant in the limits $|\vec{u}|,|\vec{v}| \rightarrow 0$ and $s,t \rightarrow \nu$. As outlined in Sec.~\ref{sec:outline}, thermal correlation functions are tempered distributions that satisfy several constraints, including the KMS condition. Due to the thermal spectral representation in Eq.~\eqref{spec_rep}, it turns out that this implies $\widetilde{D}_{\beta}(\vec{u},s)$ is \textit{also} a tempered distribution and satisfies~\cite{Bros:1996mw}
\begin{align}
\widetilde{D}_{\beta}(\vec{u},s)= d(\vec{u},s) \, e^{-\frac{\beta}{2}\sqrt{1+|\vec{u}|^{2}}},
\label{KMS_dist} 
\end{align}
where $d(\vec{u},s)$ is some other tempered distribution. Due to the $\vec{u}$-reflectional symmetry of $\widetilde{D}_{\beta}(\vec{u},s)$, it follows from Eq.~\eqref{KMS_dist} that $|\vec{u}|^{2}d(\vec{u},s)$ defines a tempered distribution in the variable $|\vec{u}|$. If one further assumes that $d(\vec{u},s)$ is \textit{regular at $|\vec{u}|=0$}, by which we mean that either $|\vec{u}|=0$ is not included in the support of $d(\vec{u},s)$, or that $|\vec{u}|d(\vec{u},s)$ is locally integrable about this point\footnote{Although this condition is not true for free fields, we expect it should hold for all non-trivial theories.}, this implies, together with Eq.~\eqref{KMS_dist} and the polynomial boundedness of the integrand coefficient, that the $|\vec{u}|$ and $|\vec{v}|$ integrals in Eq.~\eqref{shear_general_st} reduce to the integration of the tempered distributions $|\vec{u}|^{2}d(\vec{u},s)$ and $|\vec{v}|^{2}d(\vec{v},t)$ with Schwartz functions in $|\vec{u}|$ and $|\vec{v}|$, which by definition are finite. \\

\noindent
After integrating over $|\vec{u}|$ and $|\vec{v}|$, the remaining coefficient in Eq.~\eqref{shear_general_st} is a function of $s$ and $t$ that approaches a constant in the limit $s,t \rightarrow \nu$ and vanishes for $s,t \rightarrow \infty$. In order that the $s$ and $t$ integrals are finite, this therefore requires that for fixed $\vec{u}$ the thermal spectral density must satisfy the condition
\begin{align}
\int_{0}^{\infty} \! ds \, \widetilde{D}_{\beta}(\vec{u},s) < \infty.
\label{st_constr}
\end{align} 
Combining all of these results together, one is led to the following conclusion: 
\begin{displayquote}
\textit{For a theory that has a thermal spectral density $\widetilde{D}_{\beta}(\vec{u},s)$ that is regular at $|\vec{u}|=0$, has support in $\mathbb{R}^{3}\times [\nu,\infty)$ for $\nu>0$, and obeys the integral condition in Eq.~\eqref{st_constr}, if the KMS condition holds, and hence Eq.~\eqref{KMS_dist} is satisfied $\ \Longrightarrow \ \eta_{0}$ is finite.}
\end{displayquote}
This completes the proof of the condition in Eq.~\eqref{cond}. In arriving at Eq.~\eqref{eta_1}, and hence Eq.~\eqref{shear_general_st}, it was implicitly assumed that the $p_{0}\rightarrow 0$ limit and integrals could be exchanged. It turns out that the assumptions on the behaviour of $\widetilde{D}_{\beta}(\vec{u},s)$ required in order to prove Eq.~\eqref{cond} are actually sufficient to guarantee that this is valid. \\ 

\noindent 
One can now discuss the above outlined conditions on $\widetilde{D}_{\beta}(\vec{u},s)$ in the context of the specific examples of Sec.~\ref{shear_part}. For $\lambda<0$, it follows from Eq.~\eqref{D_phi4} that $\widetilde{D}_{\beta}(\vec{u},s)$ is regular at $|\vec{u}|=0$ (it vanishes), has support at $s=m>0$, and satisfies both Eqs.~\eqref{KMS_dist} and~\eqref{st_constr}, which explains why $\eta_{0}$ is finite. For $\lambda>0$, $\eta_{0}$ is divergent, and hence one (or more) of these conditions must be violated. Since $\widetilde{D}_{m,\beta}(\vec{u},s)$ is regular at $|\vec{u}|=0$ (locally integrable), has support for $s=m>0$, and satisfies Eq.~\eqref{st_constr}, it must therefore be the case that the KMS condition is violated. From the structure of the damping factor in Eq.~\eqref{D_phi4_2} one can clearly see that the exponential decay property in Eq.~\eqref{KMS_dist} does not hold, and so this confirms that the KMS condition is indeed violated in this model, as discussed in Ref.~\cite{Bros:2001zs}.

\bibliographystyle{JHEP}

\bibliography{refs1}

\end{document}